# Analysis of RF Surface Loss in a Planar 2D Qubit

Andrei Lunin, Mustafa Bal, Akshay Murthy, Shaojiang Zhu, Anna Grassellino and Alexander Romanenko

*Superconducting Quantum Materials and Systems Center,*

*Fermi National Accelerator Laboratory, Batavia, IL 60510, USA*

*Abstract*

The Josephson junction and shunt capacitor form a transmon qubit, which is the cornerstone of modern quantum computing platforms. For reliable quantum computing, it is important how long a qubit can remain in a superposition of quantum states, which is determined by the coherence time (T1). The coherence time of a qubit effectively sets the "lifetime" of usable quantum information, determining how long quantum computations can be performed before errors occur and information is lost. There are several sources of decoherence in transmon qubits, but the predominant one is generally considered to be dielectric losses in the natural oxide layer formed on the surface of the superconductor. In this paper, we present a numerical study of microwave surface losses in planar superconducting antennas of different transmon qubit designs. An asymptotic method for estimating the energy participation ratio in ultrathin films of nanometer scales is proposed, and estimates are given for the limits of achievable minimum RF losses depending on the electrical properties of the surface oxide and the interface of the qubit with the substrate material.

## Introduction

A transmon is a superconducting charged qubit that is considered to be a key element of a future quantum computer. A typical transmon design is shown in Fig. 1, where a 200 nm Nb thin film is deposited on the surface of a dielectric substrate and then, with a series of etchings, two contact pads are formed, bridged by a Josephson junction (JJ). The transmon is capacitively coupled to the readout resonator as a $\lambda/4$ segment of a coplanar waveguide (CPW). Several transmons can be placed along the feeding CPW line to create a compact experimental chip for measuring qubit parameters. Figure 2 shows the layout of such a microcircuit and its placement inside an RF box connected to two coaxial inputs.

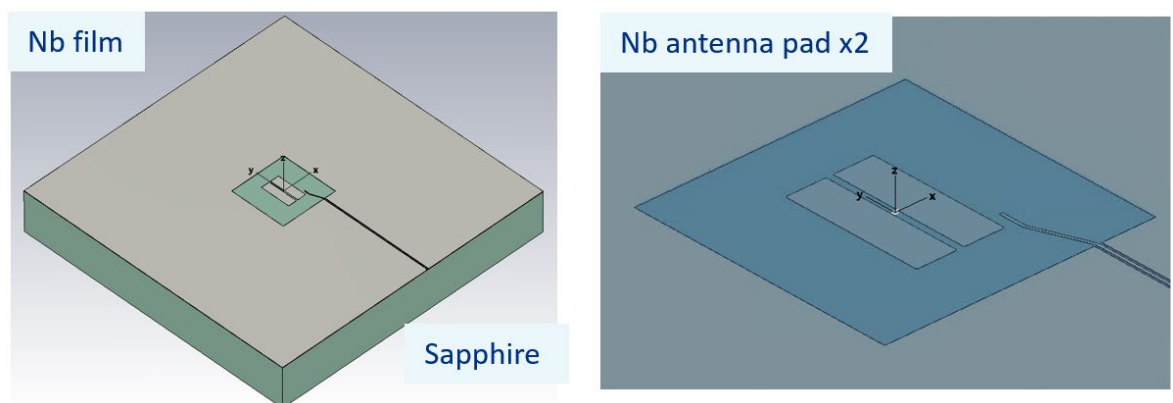

Fig. 1 Design of a superconducting charged qubit (transmon)

Since the transmon operates as a two-level system (TLS), a critical parameter is the coherence time (T1), or how long a qubit can remain in a superposition of quantum states, which is important for performing reliable quantum computations. Although there are several sources of decoherence in transmon qubits, the predominant one is generally considered to be dielectric losses in the natural oxide layer formed on the surface of the superconductor.

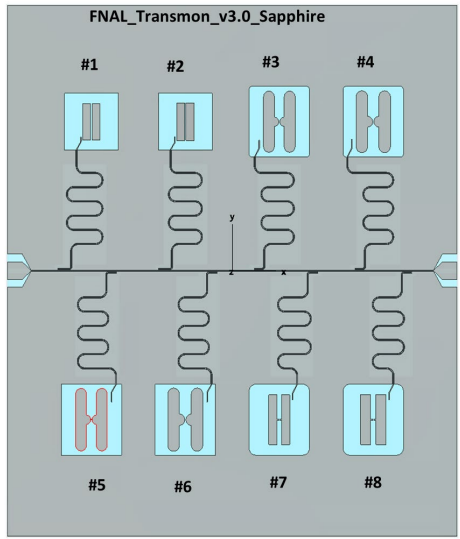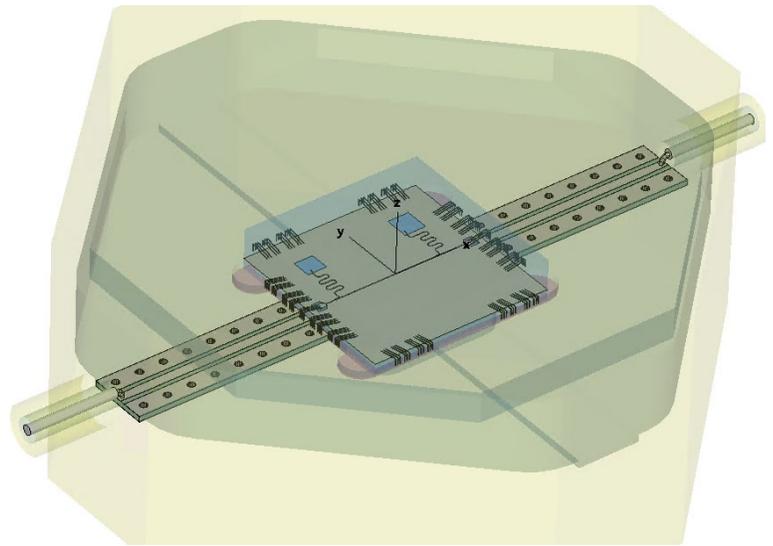

Fig. 2 Layout of the microcircuit containing eight transmons (left) and experimental setup of the chip inside a metal RF box

## Transmon Electromagnetic Modeling

The natural oxide layer formed on the surface of a metal is an amorphous dielectric, characterized by thickness and material properties, such as permittivity and loss tangent. Based on the results of SEM scanning, we proposed an idealized model of an oxide layer of a given thickness and its interface with the substrate. Figure 3 shows typical SEM images and the parameterized geometry of the Nb film, including the top and bottom rounded sidewall corners, the sidewall slope, and a possible trench formed by substrate etching.

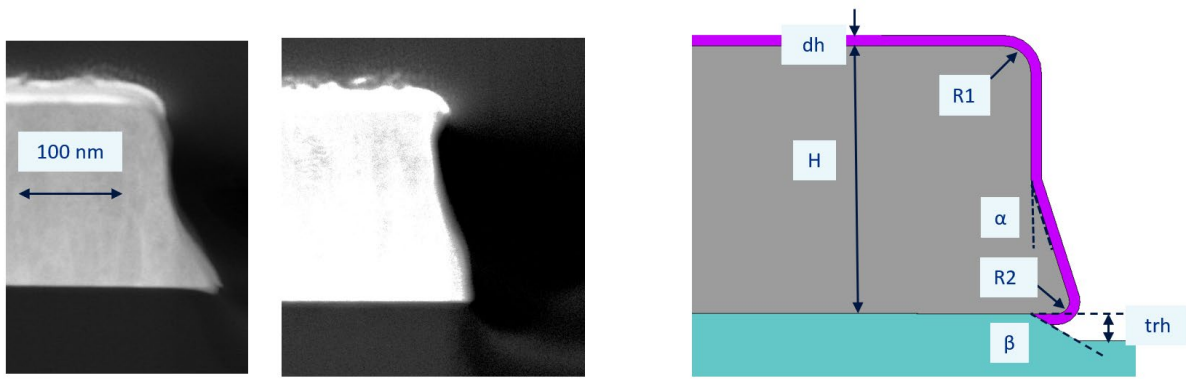

Fig .3 SEM images of the Nb film side wall (left) and corresponding parametrized geometry (right)

The two antenna pads are connected by aluminum conductors with a JJ connection between them. The JJ junction itself is presented as a lumped element, which is a rectangular surface with a given inductance. Figure 4 illustrates a model of Nb antennas and the Al conductor with two separate oxide layers on top. To improve the accuracy of the simulation, we will use the qubit symmetry planes to minimize the simulation domain, as well as the moving mesh concept.

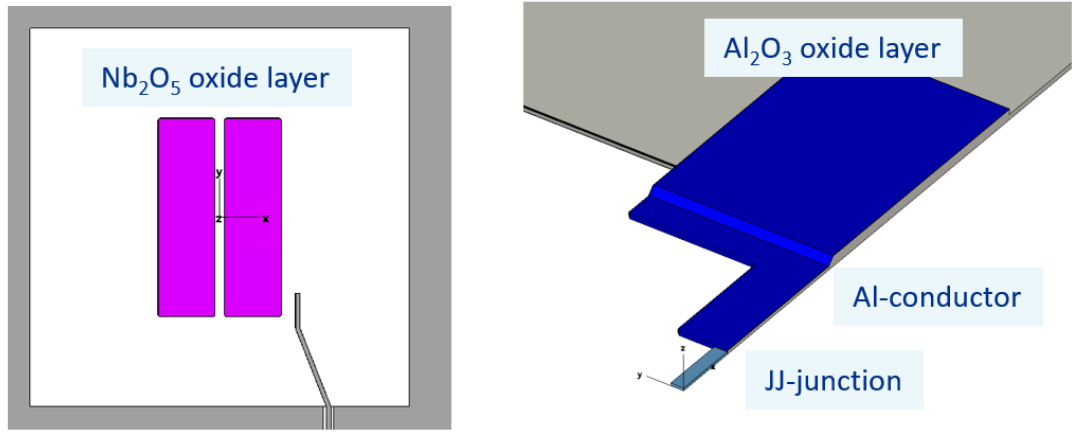

Fig. 4 Transmon 3D model of Nb antennas connected by Al conductors and JJ-junction.

The idea of a moving mesh is based on predefined virtual objects that can be reassigned to another material, such as oxide or vacuum. In this case, geometry of all objects are fixed, and the same mesh is copied and used to simulate changes in oxide thickness or substrate trench depth. Figure 5 shows such virtual objects as a conformal layered structure on the surface of a Nb film or substrate.

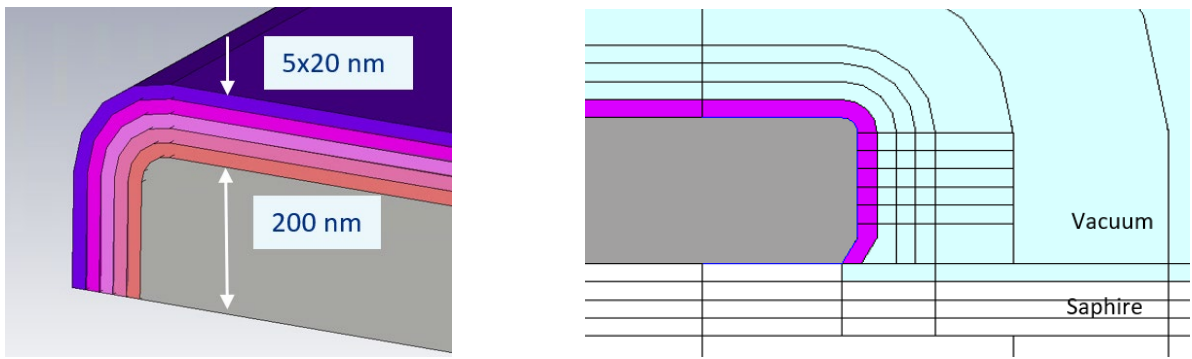

Fig. 5 Moving mesh concept, the top layers (left) represent the change in oxide thickness and the bottom layers (right) represent the substrate trench depth.

The transmon model with a lumped inductance as the JJ junction and perfect electrical boundary conditions is simulated using ANSYS HFSS code to find the qubit eigenfrequency [1]. The result is illustrated in Fig. 6, which shows the distribution of normalized electric and magnetic fields over the Nb-antenna surface and the surrounding volume. It can be seen that the electric field is mainly concentrated at the edges and corners of the antenna pads. Figure 7 shows the mesh refinement along the critical elements and a detailed view of the electric field on the top surface of the Nb film and in cross section.

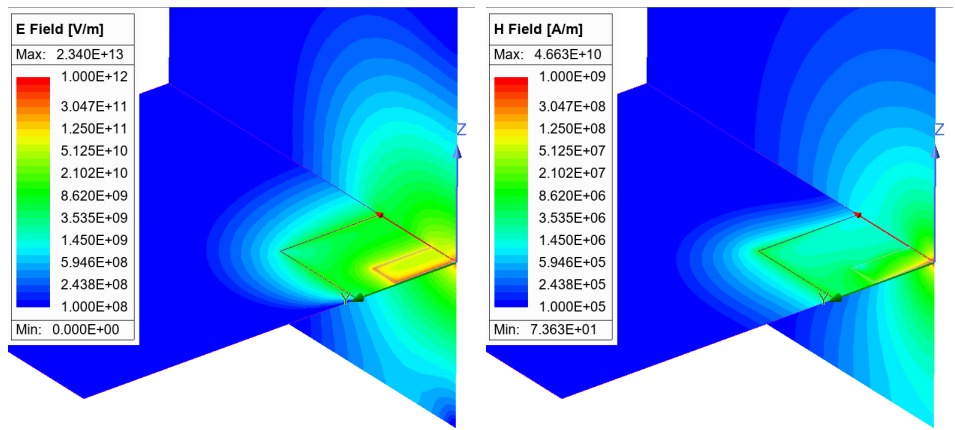

Fig. 6 Results of transmon eigenmode analysis with ANSYS HFSS, electric (left) and magnetic (right) fields normalized to 1J stored energy

The field enhancement at sharp edges is a well-known phenomenon studied in electrostatics [2]. Since the typical transmon operating frequency is around 5 GHz and its wavelengths are much larger than the nanometer scale of the sidewall rounding, the HFSS simulation is close to the electrostatic regime, where the power-law behavior of the surface charge density with distance $\rho$ from the edge can be estimated analytically for an arbitrary opening angle. We compared the calculated electric field with the electrostatic approximation for a right external corner which predicts $\rho^{-1/3}$ dependence on distance and found good agreement (see Fig. 8)

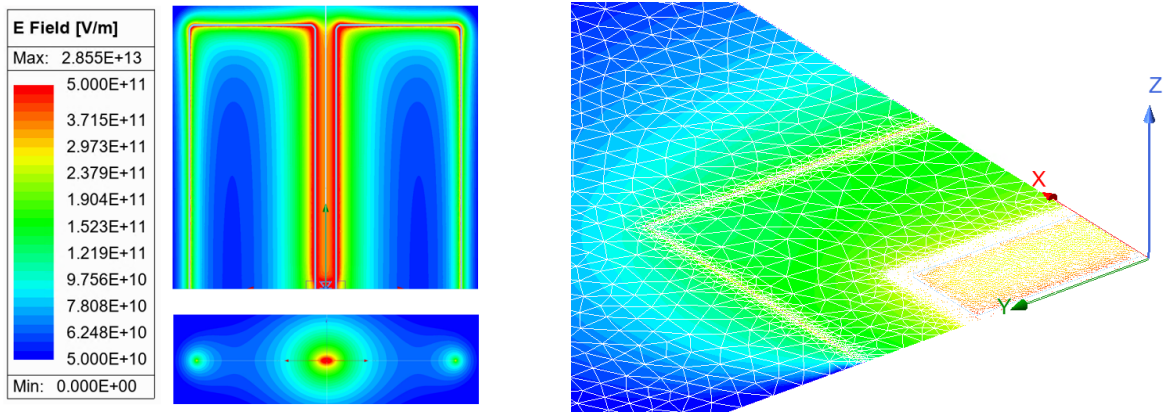

Fig. 7 Concentration of electric field at the edges and corners of the Nb-antenna pads (left) and an example mesh refinement along the antenna perimeter (right)

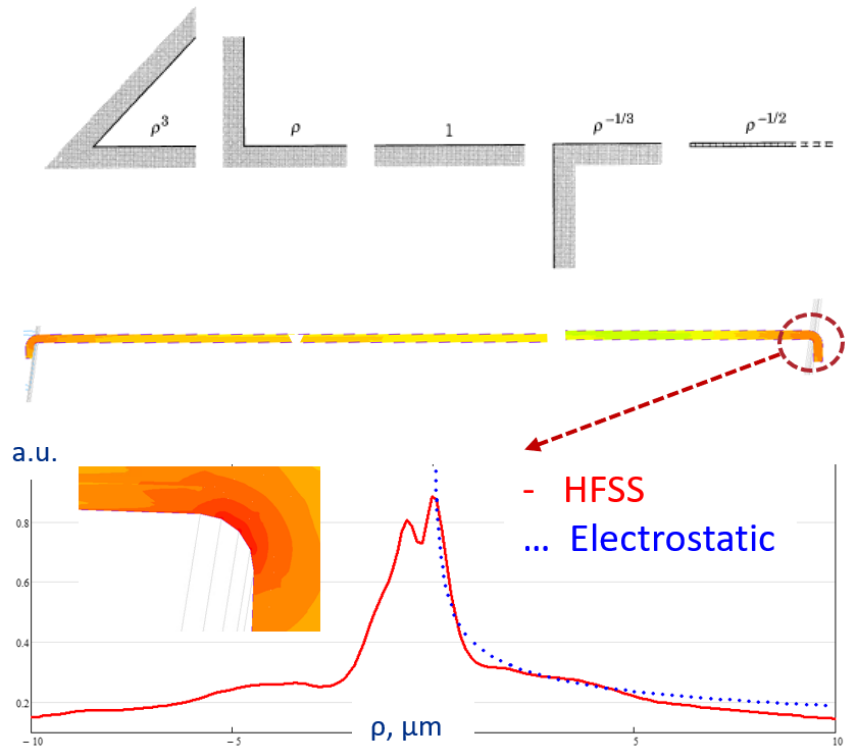

Fig. 8 Electrostatic approximations of the electric field enhancement with distance $\rho$ from a sharp edge (top) and comparison with the HFSS calculation for the case of a right external corner (bottom). The electric field is plotted for the cross-section of the oxide layer, where two peaks are the inner and outer rounding.

For the estimation of RF losses in the oxide layers we introduce the local intrinsic quality factors:

$$Q_i = \frac{\omega W_0}{P_i} = \frac{1}{epr_i * tan\_\Delta}, \qquad (1)$$

where $\omega/2\pi$ is the eigenfrequency, $W_0$ is the stored energy, $P_i = \frac{1}{2}\omega\varepsilon_0\varepsilon'' \int |E|^2 dV_i$ is the power loss, $\tan\Delta = \varepsilon''/\varepsilon'$ is the loss tangent, $\varepsilon$ is the complex permittivity of the media, $epr_i$ is the energy participation ratio (EPR) and $i$ is the number of the dielectric layer. Because in lossless resonant systems (e.g., LC circuits), the maximum stored electric and magnetic energies are equal, we can calculate the total stored energy as a balance of the electric and magnetic components as follows:

$$W_0 = W_E + (W_H + W_Q) = 2W_E = 2(W_H + W_Q), \quad (2)$$

where $W_E = \frac{1}{4}\varepsilon_0 \sum_i \int \varepsilon_i' |E|^2 dV_i$ is the total electric energy in all objects, $W_H = \frac{1}{4}\mu_0 \sum_i \int \mu_i' |H|^2 dV_i$ is the magnetic energy stored in the volume of all objects, $W_Q = \frac{1}{4} \sum_j L_j k_j \int |J_{surf}|^2 dS_j$ the magnetic energy stored in all lumped inductors, $L$ is the surface inductance (per square), $S$ is the area of the lumped inductor, $J_{surf}$ is the surface current and $k$ is the aspect ratio of the inductor area. Based on equations (1) and (2), we can calculate the EPR as the relative ratio of stored energy:

$$epr_i = \frac{\frac{1}{2}\varepsilon_0 \varepsilon_i' \int |E|^2 dV_i}{W_0}, \quad (3)$$

It is important to note here that the fields in the HFSS eigenmode solutions are by default normalized to 1 J of stored energy, but the normalization is only performed for fields in the volumes of physical objects and does not consider the energy stored in lumped elements. Thus, in this case it is necessary to explicitly calculate the stored energy for the correct normalization of the fields.

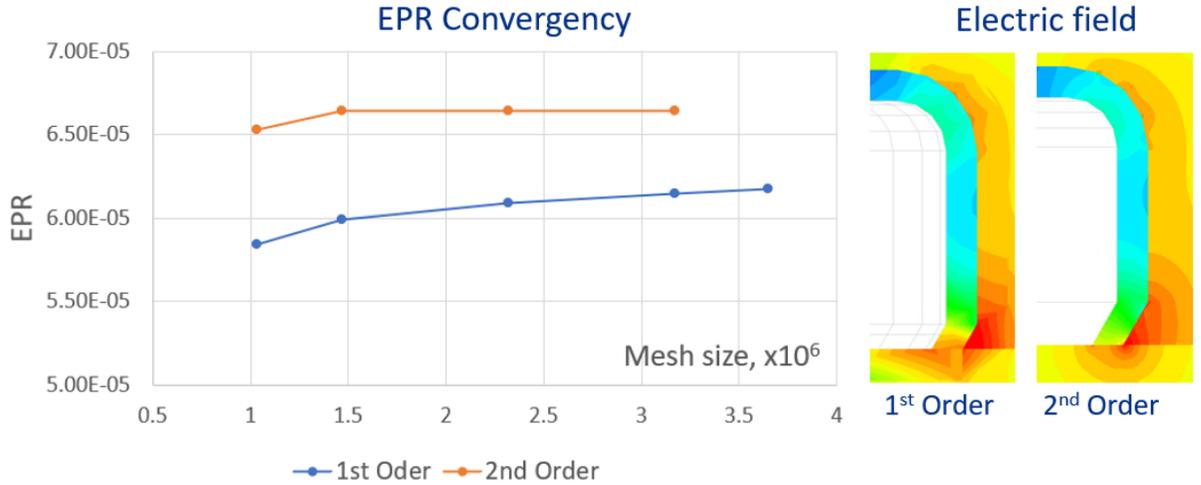

Fig .8 Study of EPR convergence depending on the mesh size and the order of the basis function (left) and examples of electric field maps for the cases of 1st and 2nd order mesh elements.

An important step to ensure the accuracy of the EPR calculations is a convergence study. The HFSS eigenmode solver uses the finite element method (FEM) to approximate the electromagnetic fields in the simulation domain. The program allows to specify the mesh elements of the 1st or 2nd order, corresponding to a linear and quadratic approximation of electromagnetic fields within each element. We ran a series of simulations, varying the order of approximation and gradually increasing the number of mesh elements. The initial mesh is seeded to ensure a denser mesh in areas of high electric field, and then the simulation goes through three iterative steps before recording the EPR value. The results are shown in Fig. 8, where it is evident that only the solution for the 2nd order mesh converges. The latter becomes clear from the electric field map near the side wall, where 2nd order elements are the preferred choice to obtain a smooth and continuous field distribution. We then performed all subsequent EPR analyses using the optimal mesh setting.

## Study of EPR Variation in Transmons with Different Antenna Geometries

One of the key components of the transmon is the antenna pads, as they are the main source of field dissipation and provide coupling to the readout resonator. In general, the shape of the antennas can be characterized by two parameters: the pad aspect ratio and the distance between the pads. Distance affects the field concentration: the closer the pads, the greater the peak electric field, and the aspect ratio determines the perimeter and the surface area of antenna, since the shorter the length, the wider the width must be to preserve the capacitance of the transmon. We studied three variants of transmons, schematically depicted in Fig. 9, with pads spacings of 20 μm, 70 μm and 150 μm, and gradually increasing pad surface [3].

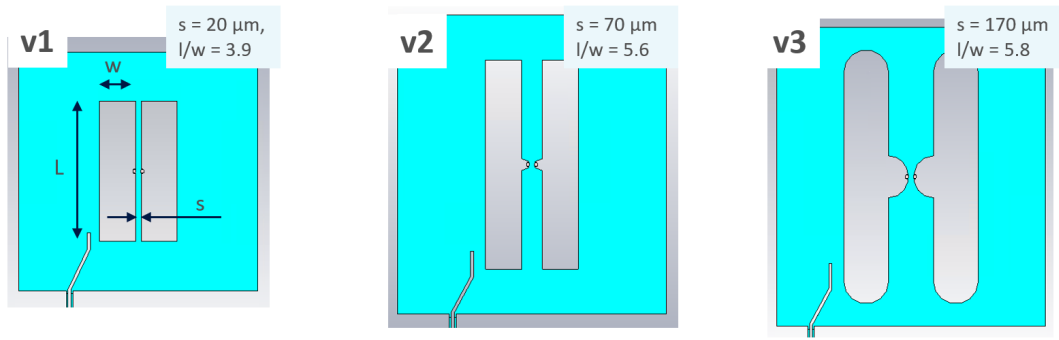

Fig. 9 Layouts of three different transmons with variations in pad spacing and pad aspect ratio.

The calculated surface electric fields are shown in Fig. 10 for all three transmons and for an oxide film thickness of 25 nm. It is noticeable that the field enhancement at the nearby edges of the antenna pads is reduced by a factor of two at large distances, and hence the same behavior can be expected for the RF surface losses. Antennas with sharp corners also cause field concentration, as in cases V1 and V2, while this effect is diminished in case V3 with rounded pads.

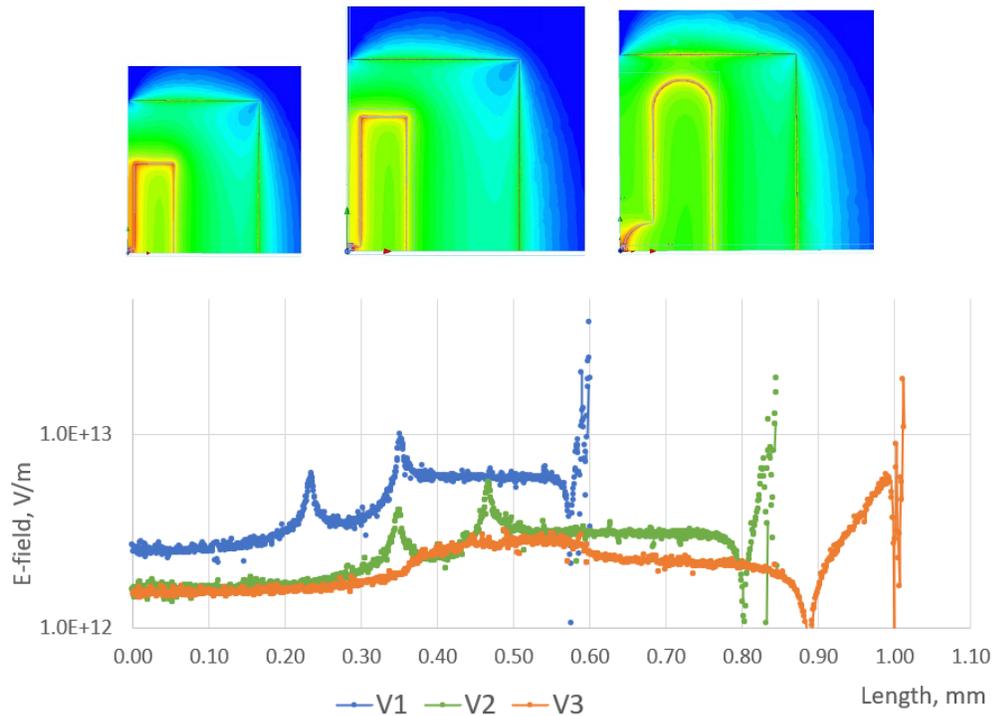

Fig. 10 Surface electric field at the top of the antenna pads (top) and at the edges along the pad perimeter (bottom)

In our calculations we used a thicker oxide layer than the typical measured thickness of about 3...5 nm. The reason is that in the case of very thin nanometer films, the HFSS mesh generator often fails to produce a mesh due to internal accuracy limitations. The remedy to this problem is to extrapolate the results obtained for thicker films to a scale of a few nanometers. One can expect a linear dependence of the EPR on the thickness, which is true for a flat metal surface when the electric field inside the dielectric layer is uniform. On the contrary, for the case of a side wall with sharp corners this assumption is not always accurate due to local field enhancements. To verify this, we investigated the dependence of the EPR on the oxide thickness up to 100 nm. The results are shown in Fig. 11, where the solid curves represent the calculated overall EPRs, and the dotted lines represent the linear extrapolation. We conclude that starting from a certain threshold of about 20...30 nm we can linearly extrapolate the EPR to the nanometer thickness range with good accuracy. For comparison, we also included the EPR of only the top surface of qubit #1 (dashed line), which is perfectly linear over the entire thickness range.

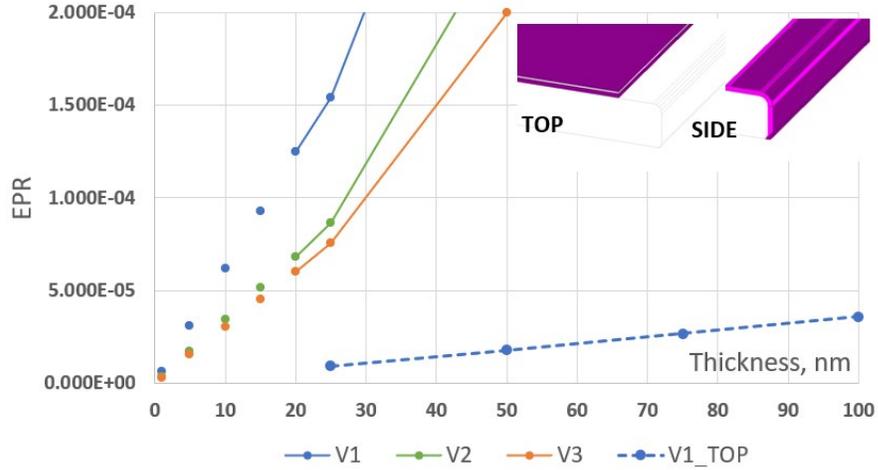

Fig. 11 Dependence of the EPR on the thickness of the oxide layer: calculated (solid) and extrapolated (dashed) EPRs for three transmon variants and the EPR of the top surface (dashed) for transmon #1. The inset images show the top and side surfaces, respectively.

Finally, we can summarize the simulation results for all three transmons in Table 1. Note that the EPRs are calculated under the condition that all dielectric materials (oxides and substrate) have a permittivity of 10, and the interface with the substrate is flat (without a trench). In addition to the EPR we also calculated G-factors of Nb antennas and the Al conductor for estimation of magnetic losses due to residual resistance.

$$G = \frac{2\omega W_0}{\oint |H|^2 dS},\qquad(4)$$

Note, that both parts of the magnetic energy (volume and lumped) should be taken into account when calculating the total stored energy. Among these three versions of transmon, V3 has the lowest calculated surface losses, as we have estimated earlier based on surface electric field analysis.

Table 1 Calculated EPR and G-factor of transmon parts

| Transmon | EPR Oxide Film | | EPR Substrate ($\varepsilon = 10$, 430 µm) | G-Factor, [Ω] | |
| --- | --- | --- | --- | --- | --- |
| | Nb-antenna ($\varepsilon = 10$, 5 nm) | Al-conductor ($\varepsilon = 10$, 4 nm) | | Nb-antenna | Al-conductor |
| V1 | 3.08E-5 | 9.0E-7 | 0.91 | 70 | 98 |
| V2 | 1.73E-5 | 5.1E-7 | 0.91 | 56 | 70 |
| V3 | 1.51E-5 | 3.8E-7 | 0.91 | 50 | 80 |

RF losses in a dielectric material depend on the product of the permittivity and the loss tangent. While these parameters can be measured for bulk material, their values in thin film containing just a few atomic layers may differ and there is a lack of reliable experimental data for such films. It is therefore useful to evaluate the EPR dependence over the entire range of expected permittivity of the oxide material. We performed a series of simulations for the V1 transmon with a 25 nm thick oxide layer, where we varied the permittivity from $\varepsilon=1$ (vacuum) to the maximum value $\varepsilon=33$ reported for amorphous niobium pentoxide [4]. The results are shown in Fig. 12 for a transmon with a flat substrate (green curve) and with an etched substrate (blue and yellow curves), as well as the theoretical limit (orange curve) of the electric field reduction in the dielectric material.

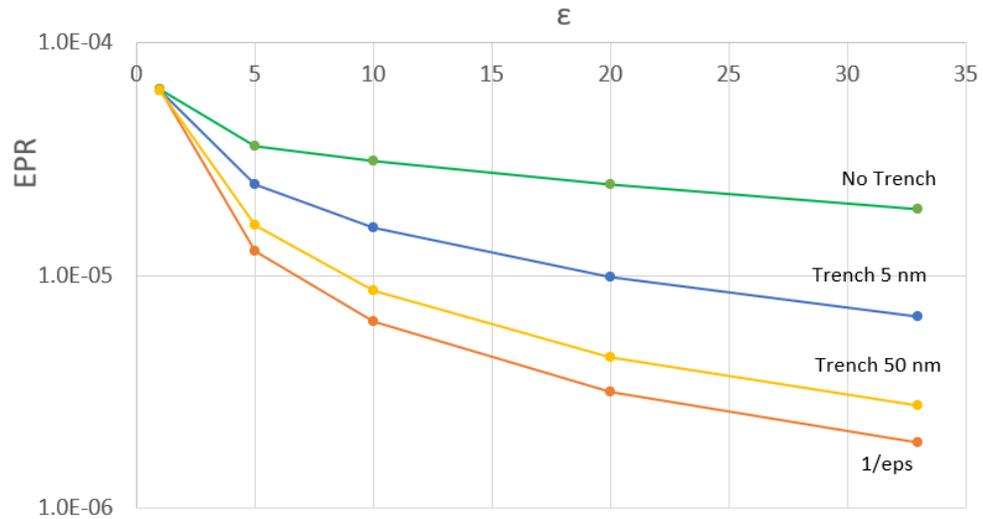

Fig. 12 Variations of EPR in Transmon V1 depending on permittivity and trench depth.

It is not surprising that the EPR of the oxide does not scale inversely with the permittivity, as predicted for a flat metal surface where only the normal component of the electric field exists. The explanation is that the electric field near sharp corners has tangential components that penetrate the oxide layer and effectively increase the stored energy and EPR. This effect is particularly evident in the lower corner of the sidewall where the oxide film interferes with the substrate. The idea is illustrated in Fig. 13, which shows the calculated field maps and the change in electric field along the perimeter of the transmon cross-section near the side wall. Note that etching the substrate helps to significantly reduce the field enhancement in the lower corner of the side wall and make the dependence of the EPR on the permittivity close to the theoretical limit.

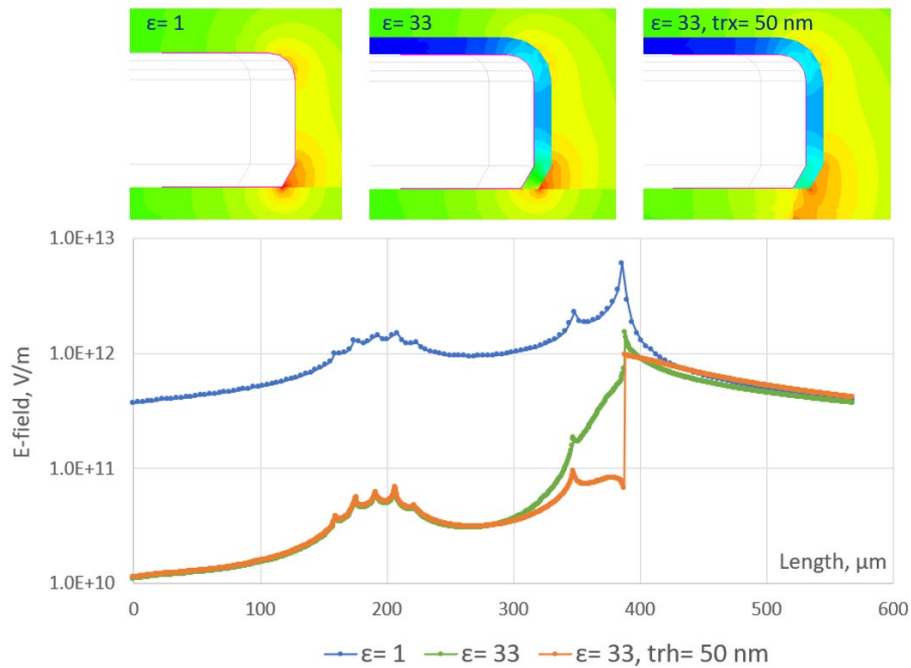

Fig. 13 Electric field maps (top) and electric field plots (bottom) along the curve (pink) going around the sidewall for three cases of oxide layers: a) $\varepsilon=1$ and flat substrate; b) $\varepsilon=33$ and flat substrate; and c) $\varepsilon=33$ and substrate etched to a depth of 50 nm.

A detailed view of the dependence of the EPR on the depth of the trench formed as a result of substrate etching is shown in Fig. 14 for surface oxides with a dielectric constant of 10 and 33. Note that the largest change in the EPR occurs in the first ten nanometers, and then both curves flatten out, since the trench depth becomes comparable to the thickness of the Nb film. Although we have not examined the etching effect for transmons other than V1, we expect the behavior to be similar, which is in good agreement with the experiment [6].

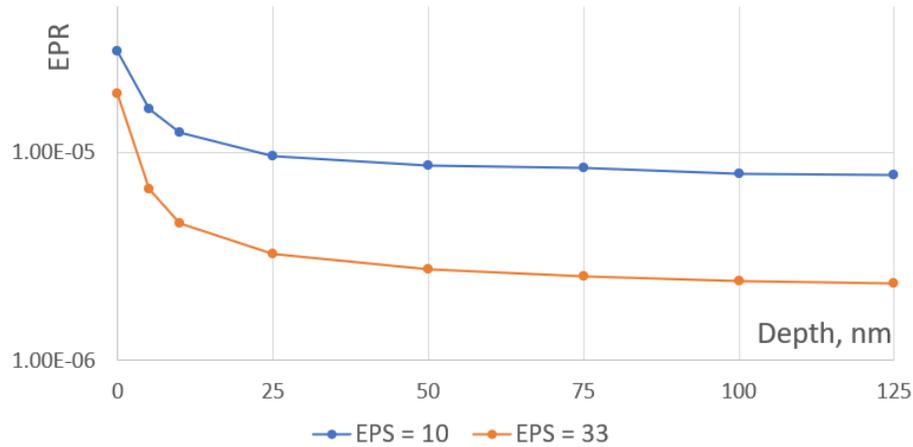

Fig. 14 Dependence of EPR on substrate etching depth for transmon V1.

## Conclusion

The main source of decoherence in a transmon qubit is the surface dielectric loss in the superconductor, caused by the presence of a natural oxide. The dielectric losses can be estimated by numerically modeling the qubit resonance and calculating the fraction of electrical energy stored in the oxide volume. The complexity of the RF modeling arises from the ultra-thin thickness of the oxide layer, which is a million times smaller than the linear dimensions of the qubit capacitive pads and makes a problematic to generate a good quality mesh. To solve this problem, we propose a method for asymptotically reducing the oxide thickness from hundreds to tens of nanometers with subsequent interpolation to units of nanometers. Finally, we applied this method to calculate the EPR in three transmon variants with pad spacings of 20 μm, 70 μm, and 150 μm, respectively. The results show the advantage of transmons with large distances between contact pads and with substrate etching, which is also confirmed in experiments.

## Acknowledgements

This material is based upon work supported by the U.S. Department of Energy, Office of Science, National Quantum Information Science Research Centers, Superconducting Quantum Materials and Systems Center (SQMS) under contract number DE-AC02-07CH11359.

## Reference

[1]   https://www.ansys.com/Products/Electronics/ANSYS-HFSS

[2]   Jackson JD, Classical Electrodynamics, 3rd ed, Wiley, 1999

[3]   S. Zhu,1, X. You, et.al., "Disentangling the Impact of Quasiparticles and Two-Level Systems on the Statistics of Superconducting Qubit Lifetime", FERMILAB-PUB-24-0577-SQMS (2025).

[4]   S. Pérez-Walton, C. Valencia-Balvín, et.al., "Electronic, dielectric, and optical properties of the b phase of niobium pentoxide and tantalum pentoxide by first-principles calculations", Phys. Status Solidi B 250 (8) (2013)

[5]   A. Murthy M. Bal, et.al., "Identifying Materials-Level Sources of Performance Variation in Superconducting Transmon Qubits", arXiv.2503.14424 (2025).